\documentstyle[prl,epsfig,aps,floats]{revtex}

\begin{document}
\draft
\def\overlay#1#2{\setbox0=\hbox{#1}\setbox1=\hbox to \wd0{\hss #2\hss}#1% \hskip
-2\wd0\copy1}
%uncomment the next line to get a two column mock up of a PRL format. SEE LINE
%BELOW THE PACS
\twocolumn[\hsize\textwidth\columnwidth\hsize\csname@twocolumnfalse\endcsname

\title{Trapping state restoration in the randomly-driven Jaynes-Cummings model
       \\ by conditional measurements}

\author{Mauro Fortunato\cite{mau}$^{(a)}$ \and Gershon Kurizki$^{(b)}$ \and
        and Wolfgang P. Schleich$^{(a)}$}
\address{(a) Abteilung f\"ur Quantenphysik, Universit\"at Ulm,
          Albert-Einstein-Allee 11, D-89069 Ulm, Germany}
\address{(b) Department of Chemical Physics, The Weizmann Institute of
             Science, Rehovot 76100, Israel}
\date{\today}
\maketitle
\begin{abstract}
We propose a scheme which can effectively restore fixed points in the
quantum dynamics of repeated Jaynes-Cummings interactions followed by
atomic state measurements, when the interaction times fluctuate randomly.
It is based on selection of superposed atomic states whose
phase correlations tend to suppress the phase fluctuations of each
separate state. One suggested realization involves the convergence
of the cavity field distribution to a single Fock state by conditional
measurements performed on two-level atoms with fluctuating velocities after
they cross the cavity. Another realization involves a trapped ion whose
internal-motional state coupling fluctuates randomly. Its motional state is
made to converge to a Fock state by conditional measurements of the internal
state of the ion.
\end{abstract}

%uncomment the next line to get a two column mock up of a PRL format. %SEE LINE
%ABOVE THE TITLE 
\vskip2pc]
\narrowtext

It is our purpose here to propose a method of controlling the evolution of a
randomly driven system by means of quantum measurements. The basic idea is
that in an ensemble of randomly phased systems it is possible to select
(project out), by appropriate quantum measurements, a {\em superposition} of
two (or more) sub-ensembles with a {\em fixed phase difference}. As a result,
the phase fluctuations of the selected superposed sub-ensembles can be
suppressed by their interference as compared to the separate fluctuations
of each sub-ensemble.

The manipulation of the system evolution by first unitarily entangling
it with another system (a ``meter''), and then post-selecting a certain
superposition of eigenstates of the meter, is known as the conditional
measurement (CM) approach~\cite{kn:cm}. This approach has been developed
for the purpose of controlling the quantum state of the cavity field by
its interactions with atoms, followed by post-selection of chosen
atomic-eigenstate superpositions~\cite{kn:cm}. The price one has to
pay for guiding the evolution by CMs is that the success probability
of every CM, which is measured by the {\em fraction} of post-selected
states among all possible measurement outcomes, is by necessity
less than unity. Hence, one must perform as many trials as implied by
the success probability until the required CM is accomplished (see
below).

Although the outlined proposal for controlling a randomly-driven system
by CMs is quite general, we shall apply it here specifically to the
fundamentally simple Jaynes-Cummings model (JCM)~\cite{kn:cm,kn:jc}
and its classical counterparts. This model describes
the interaction of a two-level dipole transition,
$|e\rangle \longleftrightarrow |g\rangle$
with a resonant quantized field mode, via the secular (rotating-wave)
interaction Hamiltonian
\begin{equation}
H_{\rm int}=g(|e\rangle\langle g|a+\mbox{\rm h.c.})\;,
\label{eq:hami}
\end{equation}
$g$ being the field-dipole coupling strength and $a$ the field
annihilation operator. This interaction can be
viewed as the resonant coupling of two quantized pendula. The merits
of this model are that it is exactly solvable (in the secular, or
rotating-wave approximation) and is experimentally realizable for
atoms crossing a high-$Q$ cavity~\cite{kn:cm,kn:jc,kn:mey,kn:miwa,kn:exp}
or for trapped ions interacting with standing laser waves~\cite{kn:ion,kn:win}.

We first discuss the cavity realization of this model.
In the interaction picture, the solution of the JCM for an atom initially
in the excited state $|e\rangle$ and the field in a superposition of many
Fock (photon-number) states,
$|\psi_{\rm F}^{(i)}\rangle = \sum_{n} c_n^{(i)} |n\rangle$,
after an interaction time $\tau$, is given by
\begin{eqnarray}
|\psi_{\rm AF}\rangle
& = & \hat{U}(\tau)|e\rangle|\psi_{\rm F}^{(i)}\rangle
\nonumber \\
& = & \sum_n c_n^{(i)}\left[\cos\theta_n|e\rangle|n\rangle
 -i\sin\theta_n|g\rangle|n+1\rangle\right] \;,
\label{eq:solution}
\end{eqnarray}
where $\hat{U}(\tau)$ is the JCM evolution operator and
$\theta_n=g\tau\sqrt{n+1}$. This solution describes a superposition of
sinusoidal energy exchanges between the ``pendula'', corresponding to
the $n$-dependent amplitudes of stimulated photon emission and
reabsorption by the atom as a function of its flight time $\tau$
through the cavity. This description neglects dissipation, which is
justifiable when the lifetimes of state $|e\rangle$ and of state
$|\psi_{\rm F}^{(i)}\rangle$ in the cavity are sufficiently
long, as discussed below. The essential feature of Eq.~(\ref{eq:solution})
is that it describes the entanglement of the field and atom states
via $\hat{U}(\tau)$.

\begin{figure}
\centerline{\hbox{\epsfig{file=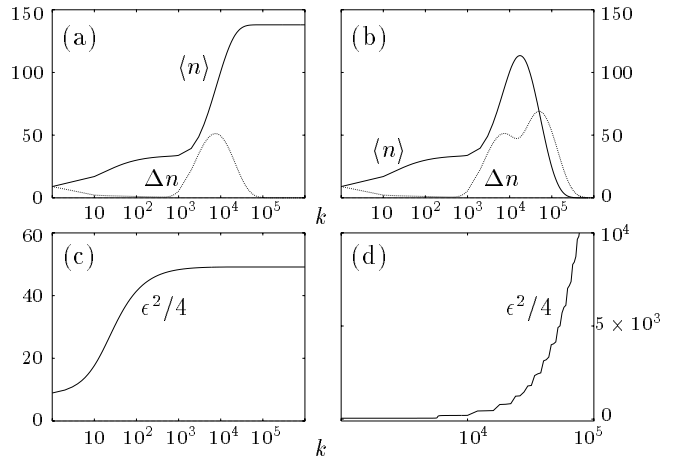,width=3.4in}}}
\vspace{0.1cm}
\caption{Sensitivity of the NSM scheme (both quantum and classical)
         to time-fluctuations. Top: behavior of $\langle n \rangle$ and
         $\Delta n$ as a function of the number of injected atoms
         in the quantum case for (a) fixed interaction times
         ($|n_t\rangle=|138\rangle$)
         and (b) tiny (1\%) fluctuations around the trapping value;
         the initial coherent state is $|\alpha\rangle=|3\rangle$.
         Bottom: evolution of $\epsilon^2/4$ (the classical counterpart
         of $\langle n\rangle$) versus number of atoms in the classical
         case for (c) fixed
         interaction times ($g\tau=2\pi/\protect\sqrt{199}$) and (d)
         tiny (1\%) fluctuations around this value; initially,
         $\epsilon^2/4=9$.}
\label{fg:one}
\end{figure}

Our study will focus on the
effect of fluctuations in the atomic velocity, and therefore in $\tau$,
on the evolution of the field state, which is driven by successive
interactions with atoms via Eq.~(\ref{eq:solution}), followed by
measurements performed on the atoms after exiting the cavity. It has
been noted~\cite{kn:jc,kn:mey,kn:meystre}
that such evolution can cause the convergence of an initial distribution
of Fock states to particular Fock states which are the ``trapping states''
of the unitary transformation~(\ref{eq:solution}), satisfying the
condition $\theta_n=q\pi$, $q$ being an integer. The originally proposed
convergence procedure relied on non-selective measurements (NSM) of the
atoms, {\it i.e.}, on ignoring the final state of each atom exiting the
cavity, and merely recording its presence~\cite{kn:meystre}. If the Fock
state distribution after the passage of $k-1$ atoms through the cavity is
$\rho^{(k-1)}_{{\rm F},nn}=|c_n^{(k-1)}|^2$,
then the entanglement with the $k$th atom according to
Eq.~(\ref{eq:solution}) followed by tracing over the final atomic states
results in the transformation
\begin{equation}
|c_n^{(k-1)}|^2 \rightarrow
|c_n^{(k-1)}|^2\cos^2\theta_n^{(k)}
+|c_{n-1}^{(k-1)}|^2\sin^2\theta_{n-1}^{(k)}\;.
\label{eq:nsm}
\end{equation}
This transformation implies that each Fock state $|n\rangle$ loses part of
its population to $|n+1\rangle$ by stimulated emission, {\em unless}
$\sin^2\theta_{n-1}^{(k)}=0$, in which case it becomes an attractor
or ``trapping'' state, whose population grows at the expense of Fock
states with {\em lower} photon numbers. Thus, a Fock-state distribution
localized {\em below} a trapping state $|n_t\rangle$ should ideally
converge to the photon-number distribution $\delta_{n,n_t}$, if all
atomic flight times $\tau_k$ are equal and exactly satisfy the above
``trapping'' (attractor) condition. However, such an attractor is merely
a {\em marginally stable} fixed point, {\it i.e.}, it is stable when
approached from below ($n < n_t$), but unstable for $n > n_t$. 
This stability is very fragile, and is destroyed even by very small
fluctuations in the interaction times around the trapping condition.
This is shown in Fig.~\ref{fg:one}(a) and (b), where we plot as a
function of the number of atoms injected the mean value and the rms
value of the photon number, in the case (a) of fixed interaction times
and (b) of tiny (1\%) time fluctuations.

These stability properties of the trapping Fock states go over smoothly
into their classical counterparts, in the limit wherein the initial
cavity field is nearly classical, allowing the replacement of the JCM
evolution [Eq.~(\ref{eq:solution})] by that of the parametrically driven
classical pendulum~\cite{kn:meystre}
\begin{equation}
{\partial\theta\over\partial(g\tau)}=\epsilon, \;\;\;\;\;
{\partial\epsilon\over\partial(g\tau)}=\sin\theta\;,
\label{eq:pendulum}
\end{equation}
where $\theta$ is the atomic polarization (the pendulum tipping
angle), $\epsilon$ is the dimensionless cavity field, and the atoms are
injected into the cavity in the excited (inverted) state. The return
map $\epsilon_{k+1}=f(\epsilon_k)$ corresponding to
Eq.~(\ref{eq:pendulum}) allows us to plot the evolution of the
field-energy distribution for fluctuating $\tau_k$. Under the assumption
of small changes in the classical field due to the interaction with one
atom, this return map can be approximated by~\cite{kn:meystre}
\begin{equation}
\epsilon_{k+1}\approx\epsilon_k+\frac{2}{\epsilon_k}\sin^2\left({\epsilon_k
g\tau_k\over2}\right)\;.
\label{eq:approxmap}
\end{equation}
The marginal stability of the fixed points, just as in the quantum domain,
makes the convergence practically impossible, and leads to {\em
intermittently chaotic} behavior, even for small $\tau_k$ fluctuations.
This is clearly shown in Fig.~\ref{fg:one}(c) and (d), where we plot
as a function of the number of atoms the classical counterpart
$\epsilon^2/4$ of the mean number of photons $\langle n\rangle$,
in the case of fixed interaction times, and of tiny (1\%)
fluctuations, respectively.

\begin{figure}
\centerline{\hbox{\epsfig{file=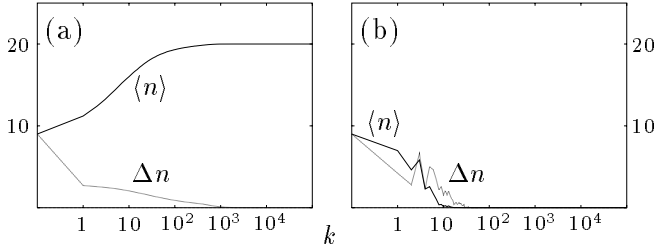,width=3.4in}}}
\vspace{0.1cm}
\caption{Evolution of $\langle n \rangle$ and $\Delta n$ versus $k$
         (number of atoms) in the scheme of elastic CMs for (a)
         small spread in the interaction times ($\Delta\tau=\Delta\tau_c/10$),
         and (b) large spread ($\Delta\tau=\Delta\tau_c$). The initial coherent
         state is $|\alpha\rangle = |3\rangle$. The random fluctuations are
         around the value $\tau_k=\pi/g\protect\sqrt{21}$, and therefore the
         trapping state (for fixed interaction time) is
         $|n_t\rangle=|20\rangle$.}
\label{fg:two}
\end{figure}

Are these fixed points irretrievable in the presence of $\tau$-fluctuations?
In order to examine this question, we consider a different kind of
field-state transformations, resulting from atomic CMs, since such
transformations can generally better guide the field evolution towards
particular states~\cite{kn:cm}. Schematically, $k$ CMs are performed
on the cavity field state if each of the $k$ atoms that have crossed the
cavity is successfully detected in the desired final state
$|\varphi^{(f)}_{k}\rangle$.
The resulting field-state transformation
$|\psi_{k-1}\rangle\rightarrow |\psi_k\rangle$
is effected by the $k$th CM according to
\begin{equation}
|\psi_k\rangle=P_k^{-1/2}\left[\langle\varphi_k^{(f)}|\hat{U}(\tau_k)
\left(|\varphi_k^{(i)}\rangle|\psi_{k-1}\rangle\right)\right]\;,
\label{eq:projection}
\end{equation}
where $\hat{U}(\tau_k)$ entangles the initial atomic state
$|\varphi_k^{(i)}\rangle$ with the field state $|\psi_{k-1}\rangle$, and
the CM success probability $P_k$ is the norm of the expression in square
brackets. The simplest choice of final atomic states
$|\varphi_k^{(f)}\rangle$ onto which the entangled state is projected
appears to be one of the energy states, $|e\rangle$ or $|g\rangle$.
For atoms injected initially in $|\varphi_k^{(i)}\rangle=|e\rangle$,
CMs consisting in post-selection of $|\varphi_k^{(f)}\rangle=|e\rangle$
or $|g\rangle$ can be referred to as the ``elastic'' and ``inelastic''
selected sub-ensembles,
respectively~\cite{kn:cm,kn:meystre}. Does the selection of one such
sub-ensemble improve the convergence of a distribution of Fock states
to a trapping state $|n_t\rangle$ when $\tau_k$ fluctuate?
As seen in Fig.~\ref{fg:two}, it does,  as long as the spread of $\tau_k$
is below the critical value $\Delta\tau_c$. This critical value can be
estimated as the difference between the trapping ($\theta_{n_t}=\pi$)
and the anti-trapping ($\theta_{n_t}=\pi/2$) interaction times. For the
``elastic'' sub-ensemble
$|\varphi_k^{(i)}\rangle=|e\rangle \rightarrow
|\varphi_k^{(f)}\rangle=|e\rangle$
this yields
$\Delta\tau_c\simeq\pi/(g\sqrt{n_t+1})-\pi/(2g\sqrt{n_t+1})$.
This estimate is fully confirmed by the numerical results in
Fig.~\ref{fg:two}: whereas for $\tau_k$-spreads
$\Delta\tau\simeq\Delta\tau_c/10$ convergence is still possible,
it fails completely for $\Delta\tau=\Delta\tau_c$, and the system
escapes its fixed points because the trapping condition is different
for each atom. 

These findings can be carried over to the quasi-classical
domain:
The selection of the ``elastic'' sub-ensemble $(|e\rangle\rightarrow|e\rangle)$
corresponds to singling out atoms that do not change, on average,
the cavity field energy by virtue of trapping.
This selection should help keep the field energy near a fixed point,
either classically or quantum mechanically, but only as long as the 
fluctuations about the fixed-point (trapping) condition are small.

\begin{figure}
\centerline{\hbox{\epsfig{file=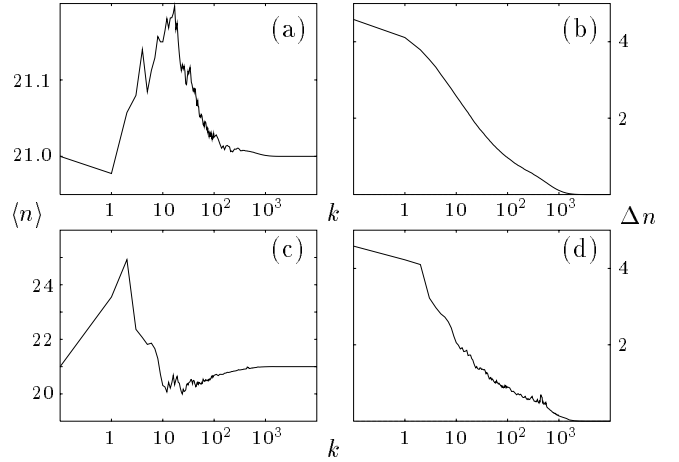,width=3.4in}}}
\vspace{0.1cm}
\caption{Stability of a Fock state---even for a large spread in
         the interaction times---in the post-selection scheme 
         $|e\rangle \rightarrow
         |e\rangle + |g\rangle$ for $|n_t\rangle=|21\rangle$.
         Top: case of small fluctuations in the
         interaction times ($\Delta\tau = \Delta\tau_c/10$): behavior of
         (a) $\langle n\rangle$ and of (b) $\Delta n$ versus $k$ (number
         of atoms). Bottom: the same for large spread in the interaction
         times ($\Delta\tau = 2\Delta\tau_c$): (c) $\langle n\rangle$ and
         (d) $\Delta n$. In all cases the initial coherent
         state is $|\alpha\rangle = |\protect\sqrt{21}\rangle$.}
\label{fg:three}
\end{figure}

However, instead of ``elastic'' or ``inelastic'' {post-se\-lec\-tion}, it
is possible to design CMs which post-select any desired superpositions
of the above sub-ensembles~\cite{kn:noi}. Such post-selections can be
experimentally realized~\cite{kn:exp} by passing the atoms after the
cavity through a classical field (Ramsey zone) setup over a time-interval
$T_k$, which rotates, say, the final state as follows
\begin{mathletters}
\label{eq:rot}
\begin{eqnarray}
|e\rangle & \longrightarrow & |\varphi^{(f)}_{k}\rangle=
\alpha^{(f)}_k|e\rangle+\beta^{(f)}_k|g\rangle\;,
\label{eq:rotstate} \\
\alpha_k^{(f)}&=&\cos(\Omega T_k/2)\;, \;\;
\beta_k^{(f)}=\sin(\Omega T_k/2)e^{i\phi_f}\;,
\label{eq:albe}
\end{eqnarray}
\end{mathletters}
$\Omega$ being the Rabi frequency of the classical field and $\phi_f$
an {\em independent} phase shift.
The corresponding CM transformation [Eq.~(\ref{eq:projection})]
$|\psi_{k-1}\rangle=\sum_nc_n^{(k-1)}|n\rangle
\rightarrow|\psi_k\rangle=\sum_nc_n^{(k)}|n\rangle$
can be written as
\begin{eqnarray}
c_n^{(k-1)} \rightarrow c_n^{(k)} & = & P_k^{-1/2}
\left[\alpha_k^{(f)*}\cos\theta_n^{(k)}c_n^{(k-1)}\right.
\nonumber \\
 & & \left.-i\beta_k^{(f)*}\sin\theta_{n-1}^{(k)}c_{n-1}^{(k-1)}\right]\;.
\label{eq:transfe}
\end{eqnarray}
If we neglect the differences between $\theta_n^{(k)}$ and $\theta_{n-1}^{(k)}$
or between $c_n^{(k-1)}$ and $c_{n-1}^{(k-1)}$, consistently with the large
$n$ limit, and use the explicit forms of $\alpha^{(f)}_k$ and
$\beta^{(f)}_k$ for a Ramsey-zone field, choosing $\phi_f=-\pi/2$,
then Eq.~(\ref{eq:transfe}) becomes
\begin{equation}
c_n^{(k)}\simeq P_k^{-1/2}\cos\left(\Omega T_k/2-g\tau_k\sqrt{n+1}\right)
c_n^{(k-1)}\; .
\label{eq:transfes}
\end{equation}
Since the atoms, whose velocities fluctuate, first cross the cavity and
then the classical field zone, $T_k$ and $\tau_k$ in Eq.~(\ref{eq:transfes})
are correlated, although random. This correlation, which results from the
{\it interference} of the post-selected sub-ensembles, yields a
{\em strong suppression} of the cosine fluctuations in Eq.~(\ref{eq:transfes}).
The consequences of this suppression are shown in Fig.~\ref{fg:three} where
we plot (for the scheme $|e\rangle \rightarrow |e\rangle + |g\rangle$)
$\langle n\rangle$ and $\Delta n$ for small and large spreads in the
interaction times: Despite the large fluctuations, the
convergence towards the desired trapping state is very good.

\begin{figure}
\centerline{\hbox{\epsfig{file=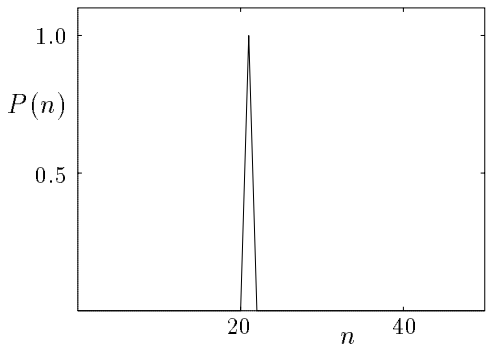,width=2.5in}}}
\vspace{0.1cm}
\caption{Final photon-number distribution after $N=2000$ atoms in
         the case of $|e\rangle\rightarrow|e\rangle+|g\rangle$ CM scheme,
         for large fluctuations in the interaction times.}
\label{fg:four}
\end{figure}

In Fig.~\ref{fg:four} the photon-number distribution
after $2000$ atoms is plotted for the case of Fig.~\ref{fg:three}(c) and (d).
This confirms that the state produced inside the cavity is indeed the Fock
state $|n\rangle=|21\rangle$: all the $P(n)$ vanish apart from one, $P(21)=1$.

The above analysis demonstrates that the superposed-subensembles CM 
allows the restoration of field
trapping states in simple dynamical evolution, as effected by  the resonant
JCM. The price one has to pay for this effect is given by the
success probability of the sequence of CMs, which amounts to post-selecting
the desired sub-ensemble and discarding all other sequences. 
For fixed interaction times
the success probabilities ${\cal P}_N$ of the whole sequence of $N$ atoms,
even though it is monotonically decreasing, converges towards
the norm of the overlap between the initial (coherent) state and the
trapping Fock state. In the case of fluctuating times, this value is
significantly smaller. Still, for the parameters of
Fig.~\ref{fg:three}(c) and (d) and for $N=110$, which suffices for the
convergence, we have ${\cal P}_N\gtrsim 10^{-3}$.

The random distribution for the interaction times $\tau_k$ we have used
in our calculation was uniform within $\Delta\tau$.
We have also made numerical runs using Gaussian distributions, with the
same rms values, obtaining similar results. The only
difference, is that in the latter
case it can happen---even though with small probability---that one atom
crosses the cavity with an interaction time which is far at the
distribution tails. This may cause the field evolution to escape
the fixed point. Different initial photon-number distributions
lead to similar results.

We notice that our results may be of value in the
task of state 
preparation~\cite{kn:cm,kn:jc,kn:mey,kn:miwa,kn:exp,kn:meystre,kn:noi}.
In particular, they may help in the generation of Fock states, since
the present scheme is quite immune even to large fluctuations
in the field-atom interaction times.

The most stringent requirement on the realization of our scheme is
that the damping time $\tau_{\rm cav}$ of the cavity be much greater
than the duration $\tau_{\rm seq}$ of the atomic
sequence multiplied by $n_t$.
A simple estimate~\cite{kn:cm} shows that for $N\simeq 10^3$ atoms with
thermal velocity ($v_{\rm at}\sim 10^3\; {\rm m/s}$) crossing a cavity
$0.5\;{\rm cm}$ long, we need a cavity quality factor $Q\gg 10^{10}$,
which is currently achievable in superconducting
microwave cavities~\cite{kn:exp}.

The above scheme can be fully implemented for a trapped cold ion
interacting with traveling- and standing-wave laser fields~\cite{kn:ion}.
The JCM evolution Eq.~(\ref{eq:solution}) then holds for the resonant
laser-induced coupling of the internal-state transition $|e\rangle
\longleftrightarrow |g\rangle$ with the vibrational state of the ion
motion~\cite{kn:ion}. The transformation of Fock states of the motion
$|n\rangle$ [Eq.~\ref{eq:nsm}] is then determined by the phases
$\theta^{(k)}_n=g\tau_k\sqrt{n+1}$, where $\tau_k$ is the duration of
the $k$-th standing-wave laser pulse. The rotation of $|e\rangle$ and
$|g\rangle$ into the superpositions given by Eqs.~(\ref{eq:rot})
is achieved by means of a traveling-wave laser pulse, whose duration
$T_k$ is {\em correlated} to that of the standing-wave pulse $\tau_k$.
Successful CMs, corresponding to Eq.~(\ref{eq:transfes}), can be performed
by detecting the fluorescence emitted by the rotated excited
state~(\ref{eq:rotstate}). These CMs can suppress the adverse effect of
laser-pulse fluctuations on the convergence of the motional state to the
trapping state $|n\rangle$, just like in the cavity realization.
The experimental difficulties of the ion trap realization~\cite{kn:win}
appear to be much less than those of the cavity realization.

To sum up, in the present scheme, the CMs force the field to evolve
towards the trapping state, in spite of the large fluctuations in the
interaction times, by using the correlations existing
between fluctuating interaction times, and measuring-device (``meter'')
times. Hence, our scheme can be considered as a novel method of restoring
fixed points in the quantum dynamics of a system which is intermittently
chaotic, even under large, random perturbations~\cite{kn:chaos,kn:cc,kn:milo}.

\acknowledgments

We acknowledge helpful and stimulating discussions with P. Stifter,
and the support of the German-Israeli Foundation.
M.F. thanks the European Community (Human Capital and Mobility
programme) for support.

\end{document}